\documentclass[12pt,epsf]{article}
\usepackage{amssymb,amsmath}
\usepackage{graphicx}

\newcommand{\be}{\begin{equation}}
\newcommand{\ee}{\end{equation}}
\newcommand{\bea}{\begin{eqnarray}}
\newcommand{\eea}{\end{eqnarray}}
\newcommand{\bear}{\begin{eqnarray}}
\newcommand{\eear}{\end{eqnarray}}
\newcommand{\beas}{\begin{eqnarray*}}
\newcommand{\eeas}{\end{eqnarray*}}
\newcommand{\ba}{\begin{array}}
\newcommand{\ea}{\end{array}}

\newcommand{\tr}{\mathrm{Tr}}

\def\identity{{\rlap{1} \hskip 1.6pt \hbox{1}}}
\newcommand{\nbox}{{\,\lower0.9pt\vbox{\hrule \hbox{\vrule height 0.2 cm \hskip 0.19 cm \vrule height 0.2 cm}\hrule}\,}}

\def\href#1#2{#2}

\begin{document}
\begin{titlepage}
\hfill
\vbox{
    \halign{#\hfil         \cr
           } 
      }  
\vspace*{20mm}
\begin{center}
{\Large \bf Evaporating Firewalls}

\vspace*{16mm}
Mark Van Raamsdonk
\vspace*{1cm}

{
Department of Physics and Astronomy,
University of British Columbia\\
6224 Agricultural Road,
Vancouver, B.C., V6T 1W9, Canada}

\vspace*{1cm}
\end{center}
\begin{abstract}

In this note, we begin by reviewing an argument (independent from 1304.6483) suggesting that large AdS black holes dual to typical high-energy pure states of a single holographic CFT must have some structure at the horizon (i.e. a fuzzball/firewall). By weakly coupling the CFT to an auxiliary system, such a black hole can be made to evaporate. In a case where the auxiliary system is a second identical CFT, it is possible (for specific initial states) that the system evolves to precisely the thermofield double state as the original black hole evaporates. In this case, the dual geometry should include the ``late-time'' part of the eternal AdS black hole spacetime which includes smooth spacetime behind the horizon of the original black hole. Thus, we can say that the firewall evaporates. This provides a specific realization of the recent ideas of Maldacena and Susskind that the existence of smooth spacetime behind the horizon of an evaporating black hole can be enabled by maximal entanglement with a Hawking radiation system (in our case the second CFT) rather than prevented by it. For initial states which are not finely-tuned to produce the thermofield double state, the question of whether a late-time infalling observer experiences a firewall translates to a question about the gravity dual of a typical high-energy state of a two-CFT system.

\end{abstract}

\end{titlepage}

\vskip 1cm

\section{Introduction}

There has been much recent debate over the fate of an infalling observer at the horizon of a black hole. Classically, the existence of solutions of Einstein's equations in which smooth spacetime continues uninterrupted past the horizon suggests that an infalling observer will notice nothing special at the horizon. Semi-classically, the absence of a large stress-energy tensor at the horizon requires a particular entanglement structure between quantum field theory modes on the two sides of the horizon. Almheiri, Marolf, Polchinski, and Sully (AMPS) have argued that such entanglement is not possible in an old black hole that has become maximally entangled with its Hawking radiation, assuming unitarity of black hole evaporation and validity of effective field theory outside the horizon \cite{Almheiri:2012rt}. Thus, they argue that an infalling observer will experience a firewall (if the assumptions hold).\footnote{See also the closely related earlier work \cite{Mathur:2009hf, Braunstein:2009my, Giddings:2011ks, Giddings:2012bm}. Since \cite{Almheiri:2012rt}, there have been many arguments for and against the firewall proposal, including \cite{lots,Almheiri:2013hfa,Avery:2013exa,Papadodimas:2012aq,Verlinde:2013uja,Verlinde:2013vja}.}

\begin{figure}
\centering
\includegraphics[width=0.6\textwidth]{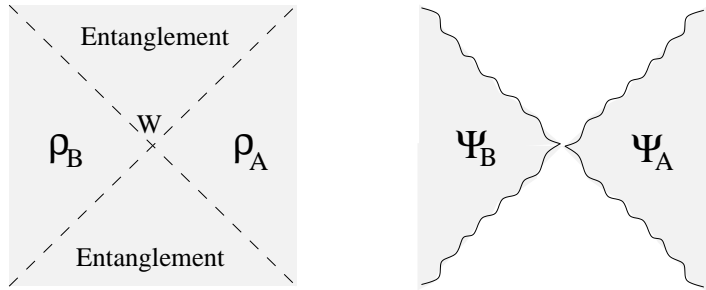}
\caption{Proposed relation between spacetime structure and entanglement structure \cite{Czech:2012be}. Density matrices for complementary sets of fundamental degrees of freedom $A$ and $B$ determine physics in the left and right wedges formed by light sheets from an extremal surface $W$. Physics in the upper and lower regions is encoded in the entanglement between $A$ and $B$. If this entanglement is removed by placing $A$ and $B$ in typical pure states in the ensembles $\rho_A$ and $\rho_B$, the right and left wedges are almost unchanged while the upper and lower regions are excised. Thus, the left and right wedges are connected by entanglement, while disentangling leaves disconnected wedges ending in a firewall.}
\label{penrose}
\end{figure}

Firewalls were also proposed in a different context in \cite{Czech:2012be}, following \cite{Israel:1976ur, Maldacena:2001kr, Freivogel:2005qh, Ryu:2006bv, VanRaamsdonk:2009ar, VanRaamsdonk:2010pw, Mathur:2010kx, Mathur:2011wg, Casini:2011kv, Czech:2012bh}. There, in the nonperturbative setting of AdS/CFT, it was argued that the existence of any spacetime at all behind a general horizon (including Rindler horizons in pure AdS) requires entanglement between the degrees of freedom associated with the region outside the horizon and some other independent degrees of freedom (see figure 1). Based on these observations, it was proposed in the context of hyperbolic AdS black holes that black hole microstates (typical pure states of a CFT on $H^d$) are dual to black holes with no spacetime past the horizon, i.e. with a lightlike singularity or firewall at the would-be horizon. As we review in section 2, very similar arguments suggest that large AdS black hole microstates described by typical high-energy states of a single CFT have firewalls (or fuzzballs). An alternative argument for this has been given recently in \cite{Almheiri:2013hfa}; our argument seems independent of this and in particular does not make reference to perturbative quantum field theory modes in the bulk.\footnote{A related discussion appears in \cite{Avery:2013exa} who argue that the gravity dual of the thermal state of a single CFT has a firewall, though we would say instead that different purifications of this thermal state correspond to different physics past the horizon which may or may not include a firewall.}

Since these large AdS microstate black holes are not maximally entangled with their Hawking radiation, they are young black holes in the sense of AMPS. Thus, their firewalls are not forced upon us by the original AMPS argument. To study ``old'' black holes in the context of AdS/CFT, and thus to evaluate the AMPS argument in a non-perturbative setting, we need to make our large AdS black holes evaporate. To achieve this, we consider in section 3 a thought experiment in which we weakly couple the CFT to an auxiliary ``radiation'' system.\footnote{A similar construction was considered in \cite{Almheiri:2013hfa}.} Specifically, we imagine that the CFT (originally in a high-energy pure state) is coupled weakly to another identical CFT, for example by placing a wire between the two spheres on which the CFTs live.

Through this weak interaction, the two-CFT system evolves from a disentangled product state to a state where the two CFTs are nearly maximally entangled (given the constraints of energy conservation). In the dual gravity picture, the interaction allows radiation from the original black hole to leak out at the AdS boundary and enter a second asymptotically AdS spacetime where it collects and eventually forms a second black hole. For a specific choice of initial state, it is possible that the entangled state that the CFTs evolve to is precisely the thermofield double state. In this case, we argue that the gravity dual to the two-CFT system includes a ``late-time'' region of the eternal AdS black hole spacetime. Thus, an observer falling into the original black hole will find smooth spacetime behind the horizon.

At least for the specific initial states that lead to the thermofield double state, our conclusions are that the black hole starts with a firewall but ends with a smooth horizon. Loosely, we could say that the black hole mellows with age and the firewall evaporates. Since the late black hole is maximally entangled with the radiation system, it is an old black hole in the sense of AMPS, and the AMPS arguments should apply. The fact that we find a smooth horizon suggests that the conclusions of AMPS can be avoided, i.e. that one of the assumptions (explicit or hidden) must be incorrect.

The thought experiment here is a specific realization of the recent ideas of Maldacena and Susskind \cite{Maldacena:2013xja}. They conjectured that the Hawking radiation can act like the second CFT in the thermofield double state; that is, the maximal entanglement between an old black hole and its Hawking radiation effectively creates a behind-the-horizon region of the spacetime. In our setup, we have simply made this more concrete by arranging that the Hawking radiation degrees of freedom {\it are} a second copy of the CFT and that the entangled state {\it is} the thermofield double state where we already know the dual spacetime. At least in this special case, we see that the spacetime behind the horizon is not forbidden by the entanglement between the black hole and its Hawking radiation; rather, it is a manifestation of this entanglement.

Ending up with the thermofield double state requires a very finely tuned initial state. This is sufficient in order to demonstrate that it is {\it possible} to end up with a smooth horizon in an old black hole, but leaves open the question of whether the firewalls proposed in section 2 generically disappear through the process of black hole evaporation. We have no definitive answer for this question, but we offer a few comments in section 4.

We conclude in section 5 with some further discussion, including comments on various related works.

\section{Fuzzballs/firewalls for black hole microstates in AdS}

In this section, we would like to ask what happens to an observer who falls into a large AdS black hole microstate described by a pure state $|\psi \rangle$ in a single CFT on $S^d$, using the ideas in \cite{Czech:2012be}.\footnote{This section is based on talks given at the 2012 Amsterdam String Workshop and at Stanford University. I am grateful to the audience members for helpful comments that have refined the argument.} We assume that this is a typical state in the thermal ensemble at some temperature.

Before reaching the horizon, the observer experiences the same Schwarzschild-AdS geometry, regardless of which black hole microstate we are in. From the CFT point of view, we can explain this by saying that the operators which tell us about the geometry outside the horizon are not sensitive to the fine details of the microstate; they may be computed using the reduced density matrix for a small fraction of the degrees of freedom. Such observables have almost exactly the same value for any typical state in the thermal ensemble (see, for example, \cite{Balasubramanian:2007qv}).

We would now like to ask what happens when the observer reaches the horizon. The AdS/Schwarzschild geometry has a standard extension to a smooth spacetime extending behind the horizon to a spacelike singularity. A simple possibility is that for any typical microstate, the observer experiences this geometry to a good approximation, at least for some substantially larger than Planck-scale distance beyond the horizon. We will now argue against this possibility.

Suppose that this hypothesis is correct. In this case, there should be some observable in the CFT that can be used to tell us about the local physics just behind the horizon.\footnote{We discuss in section 5 whether it is possible to avoid this statement.} According to the hypothesis, this observable will have approximately the same value for any typical state $|\psi \rangle$. In this case, the observable will also have the same value in the thermal state, described by a density matrix
\[
\rho = Z^{-1} \sum_i e^{-\beta E_i} |E_i \rangle \langle E_i | \;
\]
since the states $|\psi \rangle$ we are considering are typical states in this ensemble. So we could say that the thermal state $\rho$ corresponds to an AdS-Schwarzchild black hole spacetime including at least some of the region behind the horizon in the extended Schwarzchild geometry. Let us define $M_\rho$ as the geometry in common between all the typical states in the ensemble $\rho$.

The density matrix $\rho$ can also arise as a reduced density matrix starting from various pure states $|\Psi \rangle$ in theories with degrees of freedom beyond those of the single CFT. The state $|\Psi \rangle$ is a ``purification'' of the density matrix $\rho$. In some cases, the state $|\Psi \rangle$ will have a dual gravity interpretation corresponding to a classical spacetime $M$. Since knowledge of $\rho$ alone (a subset of the information in $\Psi$) allows us to deduce the existence of a classical geometry $M_\rho$, we expect that $M_\rho$ should be present as a subset of the geometry $M$ dual to any purification $\Psi$. Thus, $M_\rho$ cannot be larger than the region of spacetime common to all geometries $M$ dual to the various purifications $|\Psi \rangle$.

By considering a particular class of purifications, we will now argue that $M_\rho$ cannot be larger than the region outside the horizon of the AdS-Schwarzchild black hole. In the system we consider, the additional degrees of freedom are the degrees of freedom of a second identical CFT on $S^d$. In this system, one possible purification of $\rho$ is the thermofield double state
\[
|\Psi_{TD} \rangle = Z^{-1/2} \sum_i e^{-\beta E_i/2} |E_i \rangle \otimes |E_i \rangle \; .
\]
For this state, the dual geometry $M$ is the maximally extended Schwarzschild-AdS black hole \cite{Maldacena:2001kr}. However, there are many other possible purifications of $\rho$ in the two-CFT system. The most general purification may be written as $(U \otimes \identity) |\Psi_{TD} \rangle$, where $U$ is a unitary operator that acts only on the auxiliary CFT (corresponding to the left factor in the tensor product).

\begin{figure}
\centering
\includegraphics[width=0.4\textwidth]{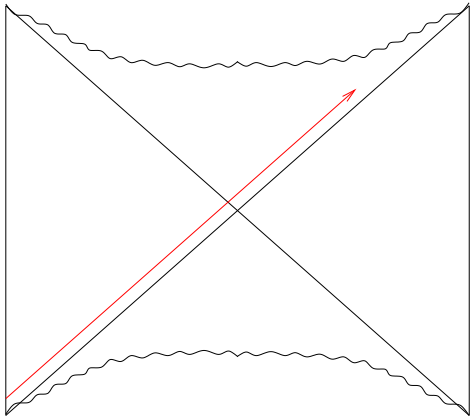}
\caption{Gravity dual of state $(U \otimes \identity)|\Psi_{TD} \rangle$ for unitary $U$ close to the identity. Perturbations originating in the second asymptotic region propagate past the horizon. For general perturbations, only the exterior of the black hole in the first asymptotic region (on the right) is unaffected.}
\label{penrose}
\end{figure}

It is easy to check that the reduced density matrix for the original CFT is $\rho$ for all these states. On the other hand, observables involving fields in the second CFT will generally differ depending on the operator $U$. From the perspective of the dual spacetime, we can think of acting with the operator $U$ as causing perturbations to the second asymptotic region.\footnote{For example, $U$ might be a local or nearly local CFT operator whose effect is to perturbs the spacetime near some boundary point.} These will propagate past the horizon and in general affect the
full spacetime everywhere except the first asymptotic region outside the horizon, as shown in figure \ref{penrose}. For a
generic state $U|\Psi \rangle$ we expect that the spacetime past the horizon of the first asymptotic
region will differ in an arbitrarily large way from the spacetime corresponding to $|\Psi_{TD} \rangle$.
Thus, the region in common to the spacetimes $M$ corresponding to the various purifications of $\rho$ in our two-CFT system is strictly the exterior of the AdS Schwarzschild black hole.

In summary, the hypothesis that gravity dual geometries for typical microstates in the thermal ensemble have a common smooth region behind the horizon suggests that the thermal state itself can be associated with a geometry $M_\rho$ including such a region. But $M_\rho$ should be a subset of any spacetime $M$ dual to a purification $|\Psi \rangle$. From the two-CFT example, it is clear that the only common region in such geometries $M$ is the exterior of the black hole. Thus, there can be no common region behind the horizon of geometries corresponding to typical microstates.

Our conclusion essentially leads to the fuzzball picture of black hole microstates \cite{Mathur:2009hf} i.e. that black hole microstates are geometries for which the region behind the black hole horizon is replaced by some structure that is different for different microstates.\footnote{A further qualitative argument for our conclusion is the following. From the CFT perspective, the observer falling into the horizon corresponds to some initially localized information being added to the CFT state (placing an observer near the boundary) and getting scrambled. It seems natural to identify the time when the observer hits the horizon with the scrambling time. At this time, the information cannot be recovered by looking at any small subset of degrees of freedom in the system. To recover the information that there was an observer, we need to use observables that are sensitive to most of the accessible degrees of freedom of the system at once. These are exactly the kind of observables we need to distinguish the individual microstates. Thus, any observables that tell us about the existence or experience of the observer after they have reached the horizon can be expected to give different results for different microstates.} For some special microstates, this structure may correspond to a smooth, weakly curved geometry, but in general, we expect that the structure is Planck-scale. The simplest interpretation is that classical spacetime simply ends at the would-be horizon, as we argued for the hyperbolic black hole microstates in \cite{Czech:2012be}.\footnote{We emphasize that it is still possible for certain special pure states to have gravity duals which contain some smooth spacetime behind a horizon. For example, we expect that state corresponding to the collapse of a null shell to form a black hole will temporarily include some smooth locally AdS spacetime inside the horizon that forms due to the collapsing shell.} In this case, an observer falling into the black hole will effectively encounter a lightlike singularity at the horizon; the fuzzball acts as a firewall.\footnote{It is not necessary that the observer burns up in a some classical way. Indeed, if it were the case that the dual of a typical microstate includes a bulk energy density that diverges in a particular way, this diverging behavior would appear also in the thermal state and thus also for the eternal black hole. We thank Don Marolf for emphasizing this point.} Our conclusion is the same as in the recent work \cite{Almheiri:2013hfa}, though the arguments here appear to be rather different.

It is important to note that the black holes we consider are not old black holes maximally entangled with their Hawking radiation. Thus, the firewalls are apparently not the ``old age'' firewalls of AMPS, but rather the ``purity'' firewalls of \cite{Czech:2012be}. To make contact with AMPS, we consider in the next section a thought experiment where we force one of these pure black holes to evaporate. Surprisingly, we find that this evaporation process can in some cases eliminate the firewall and give rise to smooth spacetime behind the horizon.

\section{Evaporating large AdS black holes}

Large black holes in Anti-de-Sitter space typically do not evaporate. They are in equilibrium with their Hawking radiation, which reaches the boundary of AdS and returns to the black hole in finite time. However, by adjusting the boundary conditions, we can allow some of the radiation to leak out to an auxiliary system (as discussed recently in \cite{Almheiri:2013hfa}). In this case, the black hole can evaporate. After some time, the black hole subsystem (which we will take to be a holographic CFT) becomes maximally entangled with the auxiliary system (which we will take to be a second holographic CFT) that contains the radiation. We can then ask what happens to an observer who falls into the black hole.

Our specific setup is the following. We start with a single CFT on $S^d$ in a typical high-energy (pure) state $|\psi_0 \rangle$ with $\langle H \rangle = E$. We consider a second identical CFT on $S^d$ that is initially in its vacuum state. We can imagine that these CFTs are living on spheres in a lab, as in the recent discussion of \cite{Maldacena:2013xja}. We now connect the two spheres by a ``wire'' so that some UV degrees of freedom in each CFT are weakly coupled.\footnote{Mathematically, we add an interaction term to the Hamiltonian. This could be of the form $H_{int} = \int_1 \int_2 {\cal O}_1(x_1) {\cal O}_2(x_2)$, where ${\cal O}_1$ and ${\cal O}_2$ have support in a small region of their respective spheres. We could also use many wires distributed around the sphere so that the interaction is approximately spherically symmetric.}

The first CFT in the state $|\psi_0 \rangle$ corresponds to a large AdS black hole microstate in a spacetime that is asymptotically global AdS. This black hole is in equilibrium with its Hawking radiation. When we attach the wire, a localized region of the AdS boundary experiences a change in boundary conditions, such that some of the radiation no longer bounces back, but is lost to an external system. Initially, there is no radiation coming in from this external system, so the black hole begins to evaporate slowly. As time passes, radiation begins to enter from the external system, and eventually the rate of outgoing radiation matches the rate of incoming radiation and the black hole (now smaller) stops evaporating.

The second CFT, initially in its vacuum state, corresponds to an empty global AdS spacetime. However, when we attach the wire, a localized region of the spatial boundary of this AdS space experiences a change in boundary conditions, and radiation begins to enter the spacetime from this region of the boundary. Initially, this radiation gives rise to a gas of gravitons in AdS space, but eventually this collapses to form a black hole that grows in size. At some point the flux of energy out through the boundary matches the flux of energy in from the boundary, so the black hole reaches an equilibrium size.

\begin{figure}
\centering
\includegraphics[width=0.7\textwidth]{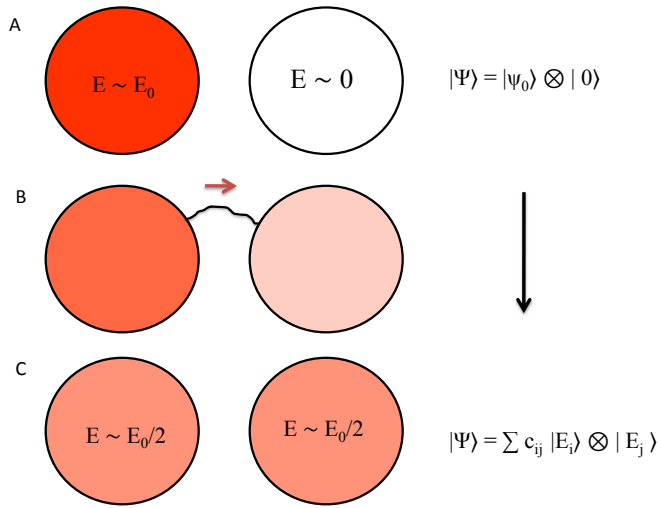}\\
\caption{CFT picture: When CFTs are weakly coupled, the product state evolves to a highly entangled state.}
\label{CFT1}
\end{figure}

In summary, the bulk picture is that after a long time, we have two black holes in different asymptotically global AdS spacetimes. Though the two CFTs start in a disentangled state $|\psi_0 \rangle \otimes |0 \rangle$, the second law of thermodynamics ensures that they will evolve to a state of the form
\be
\label{sup}
| \Psi \rangle = \sum c_{ij} |E_i \rangle_1 \otimes | E_j \rangle_2 \; .
\ee
where the entanglement entropy is close to the maximum possible value given the total energy of the system \cite{page,PSW}.
While the individual states $|E_i \rangle_1 \otimes | E_j \rangle_2$ in the superposition (\ref{sup}) correspond to disconnected spacetimes, a quantum superposition of the form (\ref{sup}) with near maximal entanglement can (at least in special cases) correspond to a spacetime in which the two asymptotic regions are connected by a smooth spacetime behind the horizons of their respective black holes\cite{Maldacena:2001kr,VanRaamsdonk:2009ar,Maldacena:2013xja}. In this case, we might expect that an observer falling into the original black hole at these late times will find some smooth spacetime behind the horizon.

For a general highly entangled state of the form (\ref{sup}), we do not know what the dual spacetime is, so it is difficult to evaluate this suggestion. However, at least for certain initial states $|\psi_0 \rangle$ and certain choices for the weak interaction Hamiltonian, we can ensure that the state of the two-CFT system after some time is precisely the thermofield double state\footnote{In fact, for any generic weak interaction between the two CFTs, the reverse time evolution of the thermofield double state will eventually (on recurrence timescales) lead to a state where almost all the energy is in the first CFT, and there is almost no entanglement between the two CFTs. Taking this to be the initial black hole state in our setup will ensure that we end up with exactly the thermofield double state at some time.}
\be
\label{TFD}
| \Psi_{TFD} \rangle = Z^{-1/2} \sum e^{- \beta E_i /2} |E_i \rangle_1 \otimes | E_i \rangle_2 \; .
\ee
In this case, the dual geometry for late times matches with the eternal black hole in AdS, and we can be certain that an observer falling into the original black hole will not see a firewall. A more detailed argument for this is presented in the following subsection.

Achieving exactly the thermofield double state clearly requires severe fine-tuning of the initial state, but a finely-tuned example suffices to show that an observer falling into a black hole that has become maximally entangled with its Hawking radiation does {\it not necessarily} encounter a firewall. Thus, the conclusions of AMPS can apparently be avoided. Apparently in this case the process of a black hole becoming maximally entangled with its Hawking radiation removes a firewall rather than creating it. As we mentioned in the introduction, our argument here is essentially the same as in the recent paper by Maldacena and Susskind \cite{Maldacena:2013xja}; the construction here simply provides a concrete example where entanglement between a black hole and its Hawking radiation can indeed lead to smooth spacetime behind the horizon.

In section 4, we return to the question of whether firewalls exist in old black holes for generic initial states.

\subsection{Gravity duals for CFTs with changing Hamiltonians}

In the previous section, we have considered a state of a two CFT system in which one CFT is originally in a typical high-energy pure state while the other CFT is in the vacuum state. At some common field theory time, the two CFTs are coupled and the state evolves to the thermofield double state. At this point, the two CFTs are decoupled again, and the entangled state evolves as usual. What can we say about the gravity dual of this system with a changing Hamiltonian?

\begin{figure}
\centering
\includegraphics[width=0.4\textwidth]{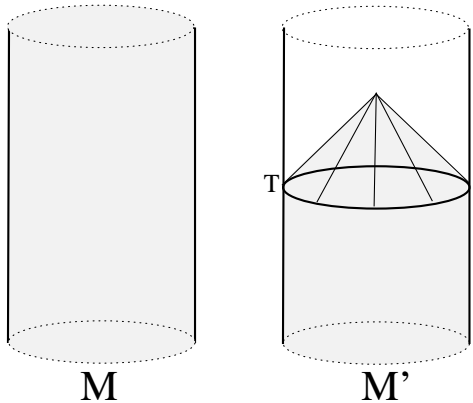}\\
\caption{Gravity dual $M'$ of a state $|\Psi'(t) \rangle$ that is the vacuum state before time $T$ but changes after time $T$ due to a perturbation of the Hamiltonian. The spacetime $M'$ matches pure global AdS in the past of the future light sheet from the boundary surface at time $T$ (shaded).}
\label{perturb}
\end{figure}

Generally speaking, suppose that in the Schrodinger picture of a holographic quantum theory with Hamiltonian $H$ we have a state $|\Psi(t) \rangle$ whose gravity dual is known to be some spacetime $M$. Now suppose we consider a Hamiltonian $H'$ that differs from $H$ after time $T$ and a state $|\Psi'(t) \rangle$ that differs from $|\Psi(t)\rangle$ only after time $T$. In the gravity dual picture, changing the Hamiltonian at time $T$ corresponds to changing the boundary conditions some boundary time. The effects of this change propagate into the bulk from the boundary surface at time $T$. Thus, the spacetime $M'$ dual to $|\Psi'(t) \rangle$ should match the spacetime $M$ in the past of the future light sheet of the boundary surface at time $T$. This is illustrated in figure \ref{perturb} for the example where $M$ is pure AdS.

Similarly, if we have a state $|\Psi(t) \rangle$ dual to $M$ and a state $|\Psi'(t) \rangle$ that differs from $|\Psi(t)\rangle$ for all times after time $T'$, the gravity dual $M'$ to $|\Psi'\rangle$ should match with $M$ at all points in the future of the past light sheet of the boundary surface at time $T'$.

\begin{figure}
\centering
\includegraphics[width=0.4\textwidth]{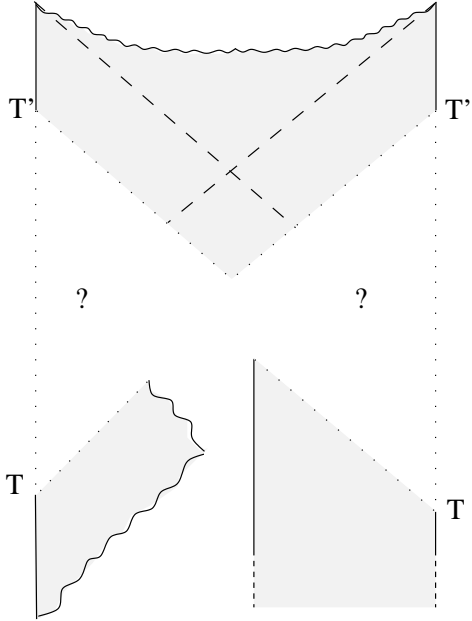}\\
\caption{Gravity dual of a state of a two-CFT system where the two CFTs are coupled between time $T$ and $T'$. Before time $T$ and after time $T'$, the CFT states are identical to states with known dual geometries. }
\label{penrose2}
\end{figure}

\begin{figure}
\centering
\includegraphics[width=0.4\textwidth]{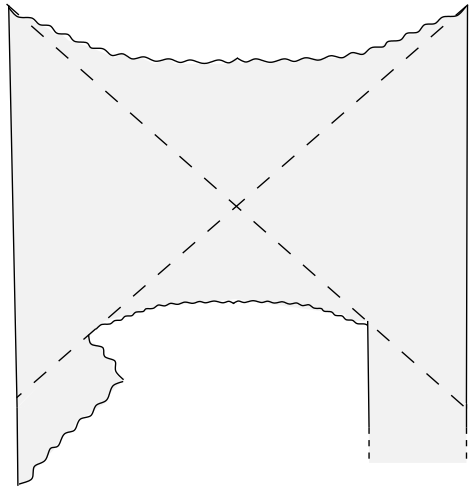}\\
\caption{Schematic of possible dual spacetime to product state of a high-energy state of one CFT with the vacuum of another CFT evolving via a weak coupling to the thermofield double state.}
\label{penrose3}
\end{figure}

Applying these requirements to the state we are interested in, we conclude that the dual spacetime to the two CFT system that is weakly coupled between times $T$ and $T'$ should include early-time portions of pure AdS and pure-state black hole geometries, and late-time portions of the eternal AdS black hole geometry, as shown in figure \ref{penrose2}. In particular, while the whole spacetime does not match the eternal black hole spacetime, the smooth spacetime region behind the horizon at late times will be present, as we require for the argument in the previous section. A schematic picture of a possible spacetime with these features is shown in figure \ref{penrose3}, though the details of this picture should not be taken too seriously.

\section{Generic case}

Ending up with the thermofield double state requires a finely tuned initial state. This is sufficient in order to demonstrate that it is {\it possible} to have a smooth horizon in an old black hole, but leaves open the question of whether a late-time observer in our setup will generically experience a firewall.

In general, the state of the system after a long time is a highly entangled state
\be
\label{sup}
| \Psi \rangle = \sum c_{ij} |E_i \rangle_1 \otimes | E_j \rangle_2 \; .
\ee
Assuming that the energy of the initial black hole is in $[E,E+dE]$ and the interaction is very weak, we can think of this final state as a typical state in the microcanonical ensemble for the two-CFT systems with total energy $E_{TOT} \in [E,E+dE]$. To answer the question of whether a late-time infalling observer in our setup experiences a firewall, we need to understand the gravity dual description of such a typical state in the two-CFT system. This is beyond the scope of the present work, but we will make a few observations.

\begin{figure}
\centering
\includegraphics[width=0.4\textwidth]{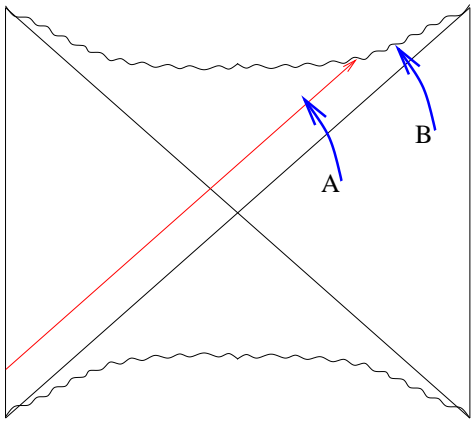}\\
\caption{Quanta falling into the black hole on one side will be observed exponentially blue-shifted by observer $A$ falling into the black hole on the other side. Observer $B$ who falls in late enough does not observe these. Back-reaction is not taken into account in the diagram, but this does not affect the conclusion.}
\label{observer}
\end{figure}

We note first that very small perturbations to the thermofield double state can lead to a firewall. Consider a perturbation that adds a small number of low-energy particles falling into the black hole on one side. As emphasized recently in \cite{Shenker:2013pqa}, an observer falling into the black hole who encounters these quanta (e.g. observer $A$ in figure \ref{observer}) will observe them to be exponentially blue-shifted. On the other hand, if the observer falls in late enough (observer B in figure \ref{observer}), the quanta will not be observed. Thus, for these small perturbations of the thermofield double state, the question of whether an infalling observer will encounter high-energy quanta behind the horizon is related to whether particles entered the horizon of the auxiliary black hole at early enough times.

The existence of not of these destructive infalling particles should be visible directly using observables in the auxiliary CFT (i.e. by knowing the density matrix of the second CFT), since they correspond to physics outside the horizon. This physics is generally believed to be encoded in ``simple'' observables that do not involve a large fraction of the degrees of freedom of the CFT. Thus, such observables can be calculated using a reduced density matrix for a small fraction of the degrees of freedom of the two-CFT system. But such a density matrix will be almost the same for any typical states of the two-CFT system. Further, it will be almost identical to the reduced density matrix for the case where the two-CFT system is in a thermal state. For this case, the single-CFT density matrix will also be thermal. Thus, we can say that for a typical state of the form (\ref{sup}), any simple observable in a single CFT will be almost identical to that observable in a thermal state of the single CFT. We conclude that for the gravity dual of a typical state two-CFT state of the form (\ref{sup}), the geometry outside the horizon in each asymptotically AdS region will be almost identical to an unperturbed black hole geometry.

Thus, we seem to have two competing effects: typicality ensures that the physics outside the horizon on each side is very close to that of the unperturbed black hole, but small perturbations on one side appear exponentially amplified to observers falling in to the black hole on the other side (if they fall in early enough). We leave a more detailed investigation of these competing effects to future work.

The preceding comments are only relevant to states that are close to the thermofield double state, so that the geometrical picture in \ref{observer} is approximately valid. For more typical states of the form (\ref{sup}), there is no reason to believe that something falling behind the horizon of the second black hole will be able to affect the experience of an observer who falls into the first black hole. Indeed, for a general state, the mutual information between any localized degrees of freedom in one CFT with localized degrees of freedom in the other CFT will be very small.\footnote{Similar comments were made in \cite{Maldacena:2013xja} and in the talk by Joe Polchinski at Strings 2013.} According to the general expectations in \cite{VanRaamsdonk:2009ar}, this means that the two asymptotic regions are very far apart; alternatively, the lack of correlations between local operators in the two CFTs suggests that there will be no spacelike geodesics connecting the two asymptotic regions. But whether or not observers from the two sides can meet behind the horizon (i.e. the two asymptotic regions are close to one another) is independent from the question of whether each observer will see smooth spacetime behind the horizon.

\section{Discussion}

If the conclusions of AMPS can be avoided, it must be that either one of the basic postulates is false\footnote{Presumably, the postulate in question is the validity of effective field theory for describing local physics outside the horizon. By assuming the validity of AdS/CFT (and further assuming that a pure state of some holographic degrees of freedom contains complete information about the corresponding dual spacetime physics) we are assuming the unitarity postulate.}, or there is a flaw in the argument. In this paper, we will not try to decide which of these is true. We simply present our observations as evidence (building on the recent discussion of Maldacena and Susskind) that ``no drama'' at the horizon is a logical possibility for old black holes. At the same time, our comments on single-CFT pure state black holes suggest that firewalls (or fuzzballs) do play a role in black hole physics, at least for large black holes in Anti-de-Sitter space. If the pure-state firewalls exist, it means that measuring the full state of the black hole in the auxiliary CFT will project the original black hole into a state with a firewall. Equivalently, measuring all the Hawking radiation should keep the firewall intact.

In the argument of section 2, it was assumed that if typical AdS black hole microstates have a common region of smooth spacetime behind the horizon, there should be some CFT operator that probes the physics in this region. Another possibility is that there is some common smooth spacetime behind the horizon of typical black hole microstate, but there is no single operator in the CFT that would tell us about it. In other words, while we can use a standard set of operators to probe the physics outside the horizon, the way to extract the physics behind the horizon is state-dependent. To avoid the firewall conclusion it must be that the state-dependent procedure applied to the individual microstates must give a different answer when applied to the mixed state describing the thermal ensemble of these microstates, or not be applicable at all. Otherwise, we would again have the conclusion that the thermal state implies some smooth spacetime behind the horizon, which we have argued cannot be correct. One example of a state-dependent construction is that of Papadodimas and Raju \cite{Papadodimas:2012aq}, which we discuss below.

In the remaining subsections, we comment on two recent proposals that would appear to be in conflict with our conclusions from section 2.

\subsubsection*{On fuzzball complementarity}

An apparent way to avoid our conclusions in section 2 is to invoke the interesting fuzzball complementarity idea of Mathur and collaborators.

In \cite{Mathur:2011wg,Mathur:2012zp,Mathur:2012jk}, Mathur and collaborators argue that while the fuzzball picture of black hole microstates is necessary to solve the information paradox, an observer with energy $E \gg T$ falling into a black hole microstate may still experience smooth spacetime behind the horizon. A key part of the argument (see, for example, \cite{Mathur_flaw}) is that the expectation value $\langle \psi|{\cal O} | \psi \rangle$ of some operator ${\cal O}$ in a typical microstate should be well approximated by the expectation value in the thermal ensemble $\tr(\rho_T {\cal O})$ which in turn is equal to the expectation value of the operator in the thermofield double state $|\Psi_{TD} \rangle$ that has smooth spacetime behind the horizon. Assuming there are operators ${\cal O}$ that can tell us about this smooth spacetime, the conclusion would be that this smooth spacetime is there, at least to a very good approximation, in the microstate.

We now argue that this conclusion is not justified. By assumption, the operator ${\cal O}$ acts only on the Hilbert space in which $|\psi \rangle$ lives. In the AdS/CFT context, this is the Hilbert space of a single CFT. But we now argue that such operators cannot possibly tell us about the spacetime behind the horizon in the thermofield double state. To see this, note that we could have replaced $|\Psi_{TD} \rangle$ in the previous paragraph by any state $ U \otimes \identity |\Psi_{TD} \rangle$, as in section 2. The expectation value of any ${\cal O}$ acting on the right-hand CFT will be identical for any $U$. But as we have argued in section 2, physics behind the horizon depends on which $U$ we choose. Thus, physics behind the horizon cannot be reconstructed using operators ${\cal O}$ acting in a single CFT.

In summary, the black hole microstates have the same logical relation to the thermofield double state as they do to any of the states $ U \otimes \identity |\Psi_{TD} \rangle$. Thus, the relation between the microstates and the thermofield double cannot be used to argue that the physics experienced by an observer in one of these microstates is the same as in the thermofield double. We have not shown that fuzzball complementarity is impossible, just that a particular argument for it does not seem justified.

\subsection{Comments on Papadodimas-Raju}

In \cite{Papadodimas:2012aq}, Papadodimas and Raju (PR) have presented an intriguing proposal for the description of the region behind the horizon of a large AdS black hole described by a pure state of a single CFT.\footnote{Another interesting recent proposal for the description of spacetime behind the horizon in a single-CFT pure state black hole appears in \cite{Verlinde:2013vja,Verlinde:2013uja}. That proposal shares some features with \cite{Papadodimas:2012aq}, though at present it is somewhat less explicit in the case of AdS.} They argue that in a typical pure state of a single CFT, a set of coarse-grained degrees of freedom will be entangled with the complementary fine-grained degrees of freedom in a way that is similar to entanglement between the two CFTs in the thermofield double state.\footnote{In \cite{Papadodimas:2012aq} the authors state that the density matrix for the coarse-grained degrees of freedom should be close to $Z^{-1} diag(e^{-\beta E_i})$, where $E_i$ are eigenvalues of a the coarse-grained Hamiltonian. This would be true if the coarse-grained degrees of freedom are very weakly coupled to the rest of the degrees of freedom, but this does not hold in a strongly interacting CFT. The more correct statement is that the density matrix for these degrees of freedom is almost identical for any typical pure state and is the same as it would be if the entire system is in a thermal state \cite{PSW}.} Based on this, they propose that the region behind the horizon for a single-CFT pure state black hole can be reconstructed from operators built from the coarse and fine grained degrees of freedom in the same way that operators are built from the the left and right CFTs to probe physics behind the eternal black hole horizon in the thermofield double state.

The PR construction seems consistent with the general picture in figure 1: if smooth spacetime exists behind the black hole horizon for a microstate black hole, we should be able to identify independent degrees of freedom in the CFT whose entanglement allows the smooth horizon. Papadodimas and Raju suggest that these are the coarse-grained and fine-grained degrees of freedom. 

If the construction of Raju and Papadodimas correctly describes physics behind the horizon of a black hole, the conclusions of section 2 must be incorrect, since PR predict a smooth spacetime behind the horizon for any typical pure state. We now argue against this possibility. While the PR construction focused on single CFT states, there is no obstacle to applying it in a case where we have a state of two CFTs entangled with one another. Thus, consider (as in section 2) the family of states of the form $(U \otimes \identity) |\Psi_{TD} \rangle$, where $|\Psi_{TD} \rangle$ is the thermofield double state. As we have argued in section 2, the physics behind the horizon of the black hole corresponding to the right-hand CFT differs depending on the operator $U$. But the density matrix of the right-hand CFT, and thus the density matrix for the coarse degrees of freedom of the right CFT, are exactly the same for all $U$. In this case, it seems that correlation functions involving ${\cal O}$ and $\tilde{\cal O}$ operators in the Papadodimas-Raju construction will not depend on $U$, incorrectly implying (according to the interpretation of PR) that the physics behind the horizon is the same independent of $U$.

\begin{figure}
\centering
\includegraphics[width=0.6\textwidth]{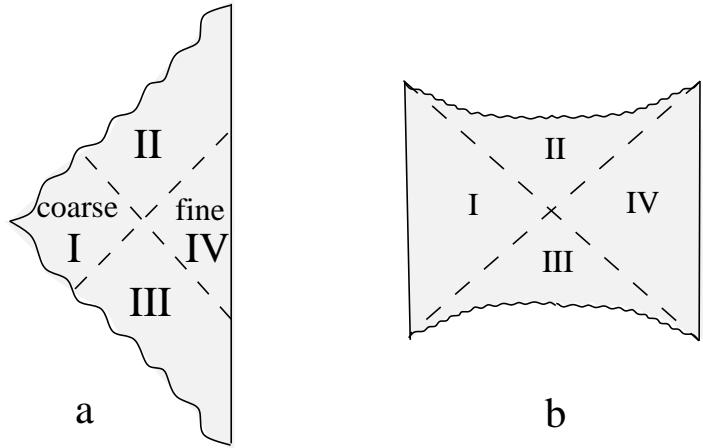}\\
\caption{a) Suggested interpretation of density matrices for coarse and fine-grained observables in a single-CFT black hole. b) Analogous regions of the AdS eternal black hole spacetime.}
\label{PR}
\end{figure}

There is an interesting alternative interpretation for the observations of Raju and Papadodimas. We suppose that our comments in section 2 are correct, so that the bulk interpretation for a typical single-CFT black hole state has the structure shown in figure \ref{PR}a. One interpretation of coarse-grained degrees of freedom would be that these are the long-wavelength degrees of freedom, conventionally taken to describe the IR physics in the bulk, i.e. physics away from the boundary. A cutoff CFT is typically understood to describe AdS with an IR cutoff at some value $R_cut$ of the radial coordinate. If we consider a CFT without a cutoff, but trace out the fine-grained degrees of freedom at some time to obtain a density matrix $\rho_{coarse}(t)$, it is therefore natural to expect that this density matrix encodes (approximately) the bulk physics on a time slice for $r<R_{cut}$, plus the physics in the bulk domain of dependence of this region (i.e. the region I in figure \ref{PR}a). Similarly, we might expect that the density matrix $\rho_{fine}(t)$ encodes the bulk physics on a time slice in a region $r > R_{cut}$ near the boundary, plus the domain of dependence of this region (i.e. the region III in figure \ref{PR}a). According to the general ideas in \cite{Papadodimas:2012aq}, reviewed in figure 1, physics in the regions II and IV should be encoded in the entanglement between the coarse and fine grained degrees of freedom. Without such entanglement, we expect that the regions II and IV would be absent.\footnote{Even in the semiclassical case, entanglement between modes in regions I and III is required to avoid a singularity at the interface of these regions that would propagate forward as a lightlike singularity on the dashed lines in figure \ref{PR}b.}

Now, we see that the regions I, II, III and IV in figure \ref{PR}a are related to each other in the same way as the corresponding regions in the eternal black hole spacetime (figure \ref{PR}b). It is therefore natural that the entanglement between coarse and fine-grained degrees of freedom in a pure state of the single CFT is qualitatively similar to the entanglement between the two CFTs in the thermofield double state.

\section*{Acknowledgments}

We are grateful to Bartek Czech, Nima Lashkari, and especially Don Marolf for helpful conversations. This research is supported in part by the Natural Sciences and Engineering Research Council of Canada and the Canada Research Chairs programme.

\end{document}